\documentclass[format=sigconf,nonacm=true]{acmart}

\renewcommand\footnotetextcopyrightpermission[1]{} % removes footnote with conference information in first column
\usepackage{epsfig}

\usepackage{amsmath}
\usepackage{algpseudocode}
\usepackage{algorithm}

\usepackage{diagbox}

\makeatletter
\def\BState{\State\hskip-\ALG@thistlm}
\makeatother

\begin{document}

%don't want date printed
\date{}

\title{Auto-tuning Distributed Stream Processing Systems using Reinforcement Learning}
%for single author (just remove % characters)
\author{Luis M. Vaquero, Felix Cuadrado
} % end author

\maketitle

% Use the following at camera-ready time to suppress page numbers.
% Comment it out when you first submit the paper for review.
%\thispagestyle{empty}

\subsection*{Abstract}
Fine tuning distributed systems is considered to be a craftsmanship, relying on intuition and experience. This becomes even more challenging when the systems need to react in near real time, as streaming engines have to do to maintain pre-agreed service quality metrics. In this article, we present an automated approach that builds on a combination of supervised and reinforcement learning methods to recommend the most appropriate lever configurations based on previous load. With this, streaming engines can be automatically tuned without requiring a human to determine the right way and proper time to deploy them. This opens the door to new configurations that are not being applied today since the complexity of managing these systems has surpassed the abilities of human experts. We show how reinforcement learning systems can find substantially better configurations in less time than their human counterparts and adapt to changing workloads.

\section{Introduction}

Processing newly generated data and reacting to changes in real-time has become a key service differentiator for companies, leading to a proliferation of distributed stream processing systems in recent years (see Apache Storm~\cite{Toshniwal2014}, Spark Streaming~\cite{zaharia2012discretized}, Twitter's Heron~\cite{Kulkarni2015}, or LinkedIn's Samza~\cite{noghabi2017samza}).

Operators must carefully tune these systems to balance competing objectives such as resource utilisation and performance (throughput or latency). Streaming workloads are also uncertain, with operators having to account for large and unpredictable load spikes during provisioning, and be on call to react to failures and service degradations. There is no principled way to fully determine a sufficiently good configuration and how to adapt it to workload changes (or available resources).

Data engineers typically try several configurations and pick the one that best matches their service level objectives (SLO)~\cite{Floratou2017}. However, individual systems have a daunting number of configuration options even for single systems. The situation is worse in distributed environments where remote interactions, networks and remote storage come into play. Finding optimal configurations is NP-hard~\cite{Sullivan2004}, making it difficult for humans to understand the impact of one configuration change, let alone the interactions between multiple ones.

This difficulty in tuning systems impacts costs, especially those related to finding highly specialised administrators. Personnel is estimated to be almost 50\% of the total ownership cost of a large-scale data system~\cite{Bara2008} and many data engineers and database administrators can spend nearly 25\% of their time on tuning~\cite{Debnath2008}. With increasing complexity, the goal of finding a working configuration in reasonable times has surpassed the abilities of humans. Indeed, Xu et al.~\cite{Xu2015} report that developers tend to ignore over 80\% of configuration options, leaving considerable optimisation potential untapped.

%Previous attempts at automatic streaming configuration tools have focused on predefined rules and building an architecture where new diagnostic heuristics can be plugged in~\cite{Floratou2017}. Machine learning (ML) has been used to find nearly optimal configurations~\cite{Nair2017} even in cases of poor sample availability~\cite{Thalheim2017}.

The configuration problem requires exploring a vast potential space, while adapting to changes to changes in order to preserve pre-established SLOs (critical in latency-sensitive applications). This context seems well suited for adaptive machine learning techniques, such as Reinforcement Learning (RL). RL systems are adopted in other domains such as self-driving cars; they take the best decision based on prior experience, while also allowing pseudo-random exploration in order to allow the system to adapt to changes. However, the adoption of RL to distributed data management system is in its early stages due to two main factors: 1) too many potential actions to explore (the famous DeepMind papers cope with tens of actions at most~\cite{silver2016}) and 2) learning which of the many monitoring metrics affect our SLO~\cite{Thalheim2017}.

%As workload or resources change, there is a need to adapt in order to preserve pre-established SLOs. This need for adaptation is much more important for latency-sensitive streaming applications. It seems that adaptive ML techniques, such as reinforcement learning (RL), similar to the ones used in self-driving cars, may be applicable to this setting. RL systems take the best decision based on prior experience, while also allowing pseudo-random exploration in order to allow the system to adapt to changes.

In this paper we address these limitations and present a system that applies RL for automatically configuring stream processing systems. We use a combination of machine learning methods that 1) identify the most relevant metrics for our SLO (processing latency), 2) select for each metric the levers that have the highest impact, and 3) discretise numeric configuration parameters into a limited set of actions.

%We present a system that selects the subset of metrics and levers that are more relevant for configuration changes to actually affect our SLO (latency in streaming). Our system can train RL models from measurements collected from previous tunings, and use the models to select the most effective configuration levers for a target objective (e.g. throughput or latency) and adapt to changing workloads. Reusing past data minimises the amount of time and resources it takes to tune a streaming system for a new application.

After training our system with a variety of workloads we show that the obtained configurations significantly improve the results obtained by human engineers. The system requires a small time to suggest configuration actions, having the additional ability to automatically adapt to changes in the streaming workload.

The remainder of this paper is organized as follows. Section~\ref{sec:back} shows the most relevant techniques used in this work. Next, in Section~\ref{sec:design} we illustrate how these techniques have been architected towards a more systematic and reproducible approach. We show how the system converges into finding a better configuration and does it in tens of minutes while being able to adapt to changing workloads in Section~\ref{sec:eval}. Section~\ref{sec:rw} highlights the most related work, while we discuss our main findings and future works in Section~\ref{sec:disc}. We finalise with a summary of the main findings of this work in Section~\ref{sec:conc}.

\section{Methodology}
\label{sec:back}

We detail in this section the techniques underpinning our automated configuration adaptation system. First, we present how we generated our training data from tens of thousands of clusters running with random configuration values. We then detail the process that automatically selects a subset of metrics and configuration levers. Finally, we present our Reinforcement Learning model for the problem of automated systems configuration.

%Thus, our requirements can be summarised as follows:
%\begin{itemize}
%\item select the metrics that best correlate with the needed SLO~\cite{Thalheim2017}
%\item identify configuration lever interdependencies~\cite{Duan2009}
%\item select the most relevant levers to drive the application SLO in the desired direction~\cite{VanAken2017}
%\item automatically cope with workload, resources and software dynamism
%\item work with numeric variables on distributed configurations (unlike~\cite{Thalheim2017}\endnote{these authors focus on binary configurations for a centralised system, different to a distributed streaming engine with continuous variables})

%\end{itemize}

\subsection{Training Data Generation}

We ran Spark Streaming clusters under various workloads and configurations (see below), in order to collect runtime performance metrics from the application as well as the infrastructure it runs on.

%Workloads
%COnfigurations
%Experiment details

We used a range of synthetic and real workloads to avoid overoptimising the model beforehand for individual scenarios. Synthetic workloads were modelled with Poisson distributions for event arrival with different $\lambda$ values, as well as with classic trapezoidal loads (ramp up, stable and ramp down period). We also used a subset of the benchmark described in~\cite{Chintapalli2016}, as well as a proprietary dataset coming from a major manufacturer of end consumer connected devices. Lists of valid values or ranges were generated for continuous variables based on the configuration of the underlying virtual machines.
%Explain how configurations were randomly generated a bit better

We have instrumented our Spark clusters with standard monitoring collection techniques\footnote{Following Spark recommendations for advanced monitoring settings, see https://spark.apache.org/docs/latest/monitoring.html\#metrics \\ OS profiling tools such as dstat, iostat, and iotop can provide fine-grained profiling on individual nodes. JVM utilities such as jstack for providing stack traces, jmap for creating heap- dumps, jstat for reporting time-series statistics. We also user perf, systemmap, gprof, systemd, as a profiling tools accounting for hardware and software events. We used a total of 90 metrics provided by these tools together with the latency and throughput of the Spark processing.}. We store per minute events forming time series of 90 metrics across all the nodes in the cluster (for each of the 80 clusters we have 9 Spark worker nodes and a driver node).

\begin{figure}[t]
\epsfxsize=8cm \epsfbox{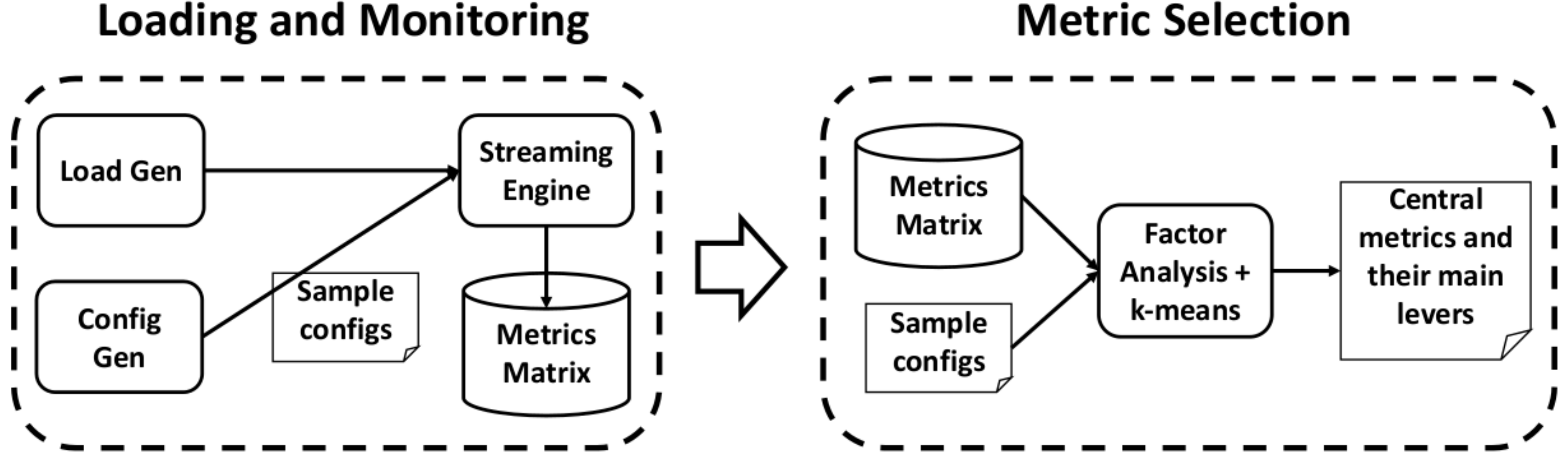}
\caption{Left: Process to generate workloads and gather monitoring data and associated lever configuration values. Right: Extraction of the most relevant metrics and their associated levers.}
\label{fig:overview}
\end{figure}

We deployed 80 Spark clusters of 10 nodes (64GB, 8 vcpus each), and ran a variety of workloads on them  (2 types of Poissonian,~\cite{Chintapalli2016}, trapezoidal, and proprietary workloads were run on 16 clusters each). Every 15 min we randomly changed the configuration of these clusters. We selected a total of 109 levers from the available ones in Spark 2.3\footnote{ https://spark.apache.org/docs/latest/configuration.html}, and changed one of them each time. Some configurations were not allowed (e.g. too low memory in the driver node) to make sure all configurations resulted in runnable conditions. In total we generated approximately $\simeq$ 100000 different configurations. The outcome of this process for each cluster is a matrix of infrastructure and application metrics along time (a matrix where one of the dimensions is time) as shown on the left hand side of Figure~\ref{fig:overview}.

\subsection{Metrics Selection}

In order to limit the processing efforts and improve clustering results, we filter out metrics showing constant trend or low variance ($\leq$ 0.002)~\cite{Thalheim2017}. This step dropped 10\% of the metrics.

As shown on the right hand side of Figure~\ref{fig:overview}, we use two classic techniques for selecting the most relevant metrics. 1) Factor Analysis (FA), which transforms the high dimensional streaming monitoring data into lower dimensional data and 2) k-means, in order to cluster this lower dimensional data into meaningful groups~\cite{VanAken2017}. The data obtained as described in the previous subsection were normalised (standardised) before doing FA. For every sample (one every 15 minutes), we took the average over 4 minutes. 

FA assumes that the information gained about the interdependencies between variables (plus an error element) can be used later to reduce the set of variables in a dataset in an aim to find independent latent variables. A factor or component is retained if the associated eigenvalue is bigger than the 95$^{th}$ percentile of the distribution of eigenvalues derived from the random data. We found that only the initial factors are significant for our Spark metrics, as most of the variability is captured by the first couple of factors.

To reconstruct missing data (e.g. due to network issues or transient failures), we use 3$^{rd}$ order spline interpolation to minimise distortion to the characteristics of time series of metrics~\cite{Liu2009}.

The FA algorithm takes as input a matrix $X$ with metrics as rows and lever values as columns, thus entry $X_{i, j}$ is the value of metric $i$ on lever value $j$. FA returns a lower dimension matrix $U$, with metrics as rows and factors as columns, and the entry $U_{i,j}$ is the coefficient of metric $i$ in factor $j$. The metrics are scatter-plotted using elements of the $i^{th}$ row of $U$ as coordinates for metric $i$.

The results of the FA yield coefficients for metrics in each of the top two factors. Closely correlated metrics can then be pruned and the remaining metrics are then clustered using the factor coefficients as coordinates.

As metrics will be close together if they have similar coefficients in $U$, we clustered the metrics in $U$ via k-means, using each metrics row of $U$ as its coordinates. We keep a single metric for each cluster, namely, the one closest to the centre of the cluster. We iterated over several k values and took the number that minimised the cost function (minimum distances between data points and their cluster centre)~\cite{VanAken2017}.

\begin{figure}[t]
\center
\epsfxsize=8cm \epsfbox{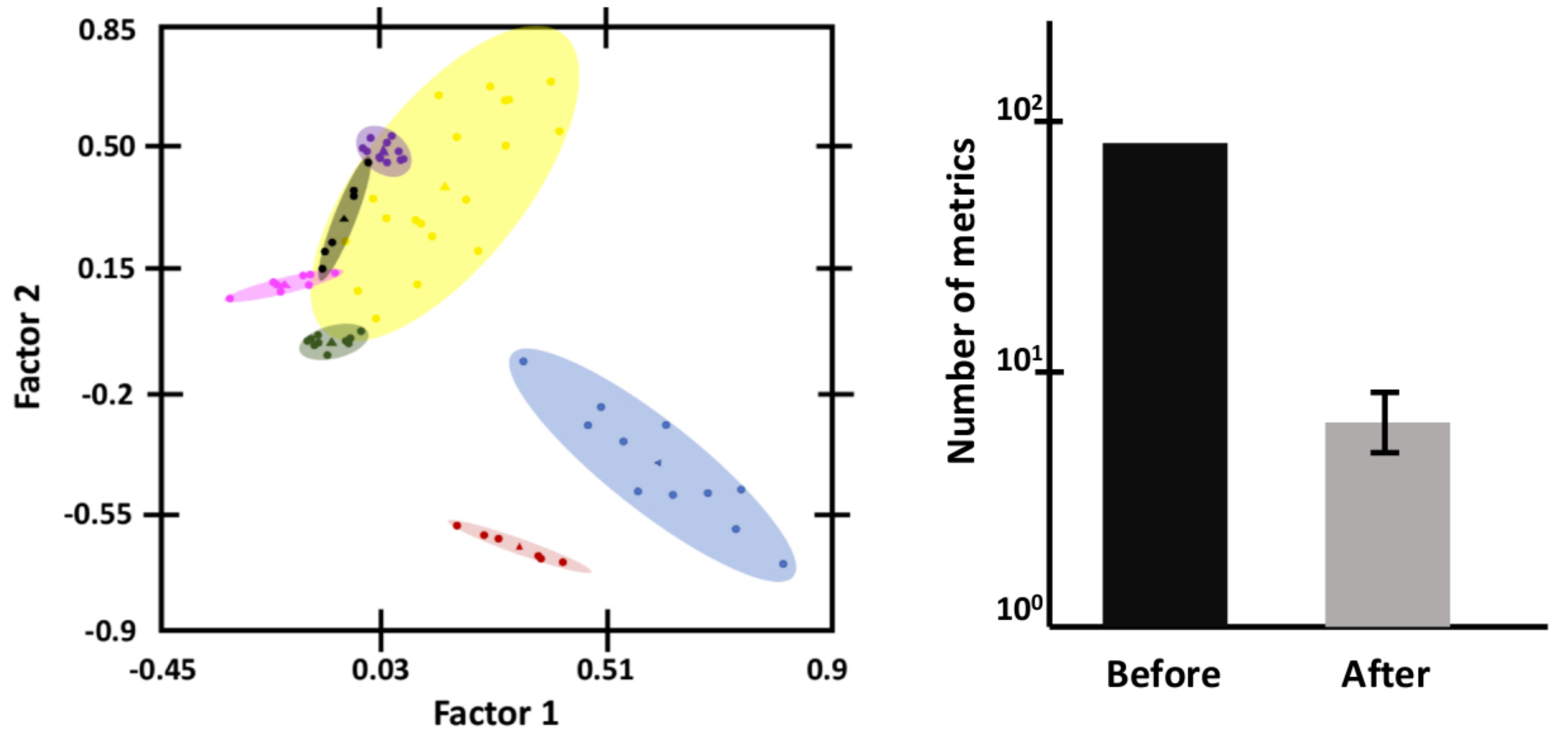}
\caption{Clusters of metrics resulting from the FA + k- means analysis}
\label{fig:fa}
\end{figure}

Figure~\ref{fig:fa} (left) shows a two-dimensional projection of the scatter-plot and the metric clusters. This process identifies a total of 7 clusters, which correspond to distinct aspects of a system's performance. Our results show some expected relationships (cache performance counters are in the same group as metrics related to JVM performance, like garbage collection). We can also see that some of these metrics are well-organised (e.g. metrics related to overall input-output performance are close to memory metrics but they fall under different categories). From the original 90 metrics, we were able to reduce the number of metrics by 92\% (this result varies slightly for different runs of the same data due to the random initialisation of the clusters, shown in Figure~\ref{fig:fa} (right)). The FA plus clustering analysis is run separately in two batches: 1) the Spark driver node and 2) all the Spark worker nodes, in order to assess adequately the metrics that are exclusive to the driver (e.g. Spark driver memory).

\subsection{Ranking Most Actionable Metrics per Lever}
\begin{figure}[t]
\epsfxsize=8cm \epsfbox{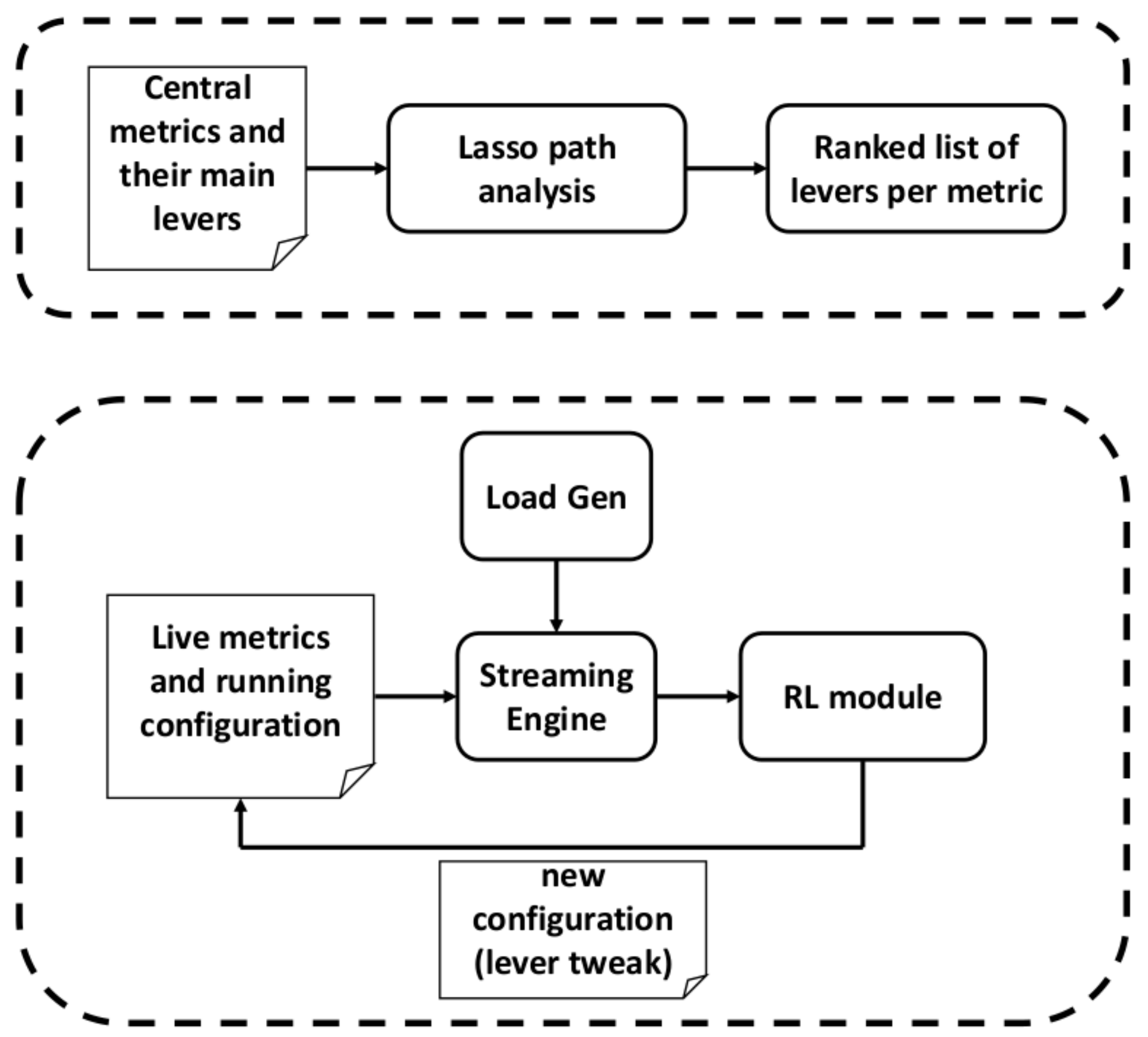}
\caption{Top: election of levers with strongest correlation with performance. Bottom: Reinforcement learning configuration feedback loop}
\label{fig:prun}
\end{figure}

%Streaming engines can have hundreds of levers, but some may be redundant (have positive and negative correlations) in their effect on the measured performance. For example, reducing the amount of memory allocated for the streaming engine driver is likely to degrade the systems overall latency.
Having reduced the metric space, we then try to find the subset of configuration levers with the highest impact on the target objective function. We use a feature selection technique for linear regression (including polynomial features)~\cite{Tibshirani2011}. The process can be observed in Figure~\ref{fig:prun} (top).

We represented the configuration levers to be tuned with the set of independent variables $R$, whereas the linear regression dependent variables $Y$ represented the preselected metrics. We convert categorical valued levers into continuous valued variables by numbering the categories. These variables are then normalised (value minus mean divided by standard deviation). The Lasso adds a L1 penalty equal to a constant $\lambda$ times the sum of absolute weights to the cost function. Because each non-zero weight contributes to the penalty term, Lasso effectively shrinks some weights and forces others to zero, thus automatically selecting features (non-zero weights) and discarding others (zero weights)~\cite{VanAken2017}.

As indicated by~\cite{Hastie2001}, we initially start with all weights set to zero (no features selected). We then decrease the penalty in small increments, recompute the regression, and track what features are added back to the model at each step. The order in which the levers first appear in the regression determines how much of an impact they have on the target metric~\cite{VanAken2017}.

In our experiments, each invocation of Lasso takes 30 min and consumes 20 GB of memory. The Lasso path algorithm guarantees that the selected levers are ordered by the strength of statistical evidence and that they are relevant (assuming that the data are well approximated using a linear model).

\subsection{Automated Tuning}

\subsubsection{Dynamic Lever Discretisation}

Before the configurator can perform any modification on any of the lever, it needs to discretise the metrics. We follow a process for dynamically categorising continuous variables as described in~\cite{Subiros2015}. Briefly, each lever corresponding to a continuous variable is manually marked with the min and max values from the samples data. The initial bin size is set to: $\delta=\frac{|max - min|}{10}$. If the RL configurator assigns the maximum bin for a number of times, the max is increased by one bin (a new bin is added so that $new-max = max + \delta $). The bin size is dynamically updated as follows: if the same bin is assigned a configurable number of times, then the bin size is halved. If this happened for the first time, then we would have 20 bins after this initial halving. The algorithm can also merge bins as described in~\cite{Subiros2015}.

The central value of the bin is taken as value for the configuration parameter. We also add a smaller ridge term for each configuration the RL configurator selects. The ridge term adds/substracts a small value to the central value of the bin. This is helpful for ``noisy'' cloud environments. This ridge factor means that the configuration chosen by the configurator for some of the levers is modulated to the top or the bottom of the discretisation (binning) interval.

\subsubsection{Reinforcement Learning Configurator}

At this point the system has (1) the set of non-redundant metrics; (2) the set of most impactful configuration levers; and (3) a mechanism to dynamically discretise continuous levers. The configurator now needs to:

\begin{itemize}
  \item learn a mapping from non-redundant metrics to impactful configuration levers (this mapping has been massively pruned by Lasso)
  \item select a lever, and decide whether to increase or decrease the value
\end{itemize}

%A RL agent (spark configurator) interacts with an environment (spark stream configuration and metrics)~\cite{Sutton1998}.

At each time step $t$, the agent observes some state $s_{t}$ (values for 109 levers and 90 metrics), and is asked to choose an action $a_{t}$ (change in the value of one of the levers) that triggers a state transition to $s_{t+1}$ and the configurator receives reward $r_{t}$. We used a delay-dependent reward (see Section~\ref{sec:design} below).

Both, transitions and rewards are stochastic Markov processes (probabilities and rewards depend only on the state of the environment $s_{t}$ and the action taken by the configurator $a_{t}$). The configurator has no \textit{a priori} knowledge of the state the system will transition to, neither consequentely about the reward it will obtain. The interaction with the configured system and collection of rewards is what drives learning. The goal of learning is to maximise the $E[ \sum_{t=0}^{\infty} \gamma^{t} * r_{t} ]$ where $\gamma \in (0, 1]$ is a factor discounting future rewards.

The configurator picks actions based on a policy, defined as a probability distribution over actions: $\pi : \pi(s,a) \rightarrow [0,1]$ where $\pi(s,a)$ is the probability of picking action $a$ (lever value) in state $s$.

On every state, the configurator can choose from several possible actions and it will take one or another based on prior experience (rewards). This is similar to a big lookup table, represented by a state-action value function $Q(s,a)$.

$Q(s,a) = r + \gamma * max_{a'} Q(s',a')$

The equation in essence means that the value of a state depends on the immediate reward $r$ and a discounted reward modulated by $\gamma$.

As there are too many states (109 possible levers can be acted on by increasing or decreasing them and 90 metrics with continuous values), it is common to use function approximators for $Q(s,a)$~\cite{Mao2016}. A function approximator,  $\theta$, has a manageable number of adjustable parameters; we refer to these as the policy parameters and represent the policy as $\pi(s,a)_{\theta}$. Note that we use a stochastic policy (uniform random) to select among the set of filtered levers.

Deep Neural Networks (DNNs) have recently been used successfully as function approximators to solve large-scale RL tasks~\cite{silver2016}. An advantage of DNNs is that they do not need hand-crafted features.

While the goal of Q Learning is to approximate the Q function and use it to infer the optimal policy $\pi^{*}$ (i.e. $argmax_{a} Q(s,a)$), Policy Gradients use a neural network (or other function approximators) to directly model the action probabilities.

Each time the configurator interacts with the environment (generating a data point $<s,a,r,s'>$), the neural network parameters $\theta$ are tuned using gradient descent so that ``good'' tuning to the configuration levers will be more likely used in the future. The gradient of the objective function above is:

$\nabla_{\theta} E_{\pi_{\theta}}[ \sum_{t=0}^{\infty} \gamma^{t} * r_{t} ] = E_{\pi_{\theta}} [\nabla_{\theta} log(\pi_{\theta}(s,a)) Q^{\pi_{\theta}}(s,a)]$

where $Q^{\pi_{\theta}}(s,a)$ is the expected cumulative discounted reward from (deterministically) choosing action $a$ in state $s$, and subsequently following policy $\pi_{\theta}$.

Using Monte Carlo Methods~\cite{hastings70}, the agent samples multiple trajectories and uses the empirically computed cumulative discounted reward, $v_{t}$, as an unbiased estimate of $Q^{\pi_{\theta}}(s_{t}, a_{t})$. It then updates the policy parameters ($\theta$) via gradient descent:

$\theta \leftarrow \theta + \alpha \sum_{t} * \nabla_{\theta} log(\pi_{\theta}(s_{t},a_{t})) v_{t}$

where $\alpha$ is the step size. We implemented the well-known REINFORCE algorithm~\cite{Sutton1999}. The direction of $\nabla_{\theta} log(\pi_{\theta}(s_{t},a_{t}))$ gives how to change the policy parameters in order to increase $\pi_{\theta}(s_{t},a_{t})$, the probability of action $a_{t}$ at state $s_{t}$. The size of the step depends on the size of the return $v_{t}$, this means that actions that empirically lead to better returns are reinforced. In order to decrease the variance of gradient estimates based on a few local samples we subtract a baseline value from each return $v_{t}$.

The most relevant levers are preferentially used by our RL algorithm (the top lever is used $f$ \% of the time), but the other levers will also be used occasionally ($1 - f$) to keep a good trade-off between exploration and exploitation (see next section).

\section{Design}
\label{sec:design}

We represent the state of the system (the current monitoring metrics and the key actionable levers) as distinct images (one for each of the monitoring metrics) and another for showing the discretised configuration values.

As in~\cite{Mao2016}, we keep a grid per metric, where each cell represents a node in the cluster. There is a matrix for each resource showing the average utilisation of the resources during the running of that configuration.

\begin{figure}[t]
\epsfxsize=8cm \epsfbox{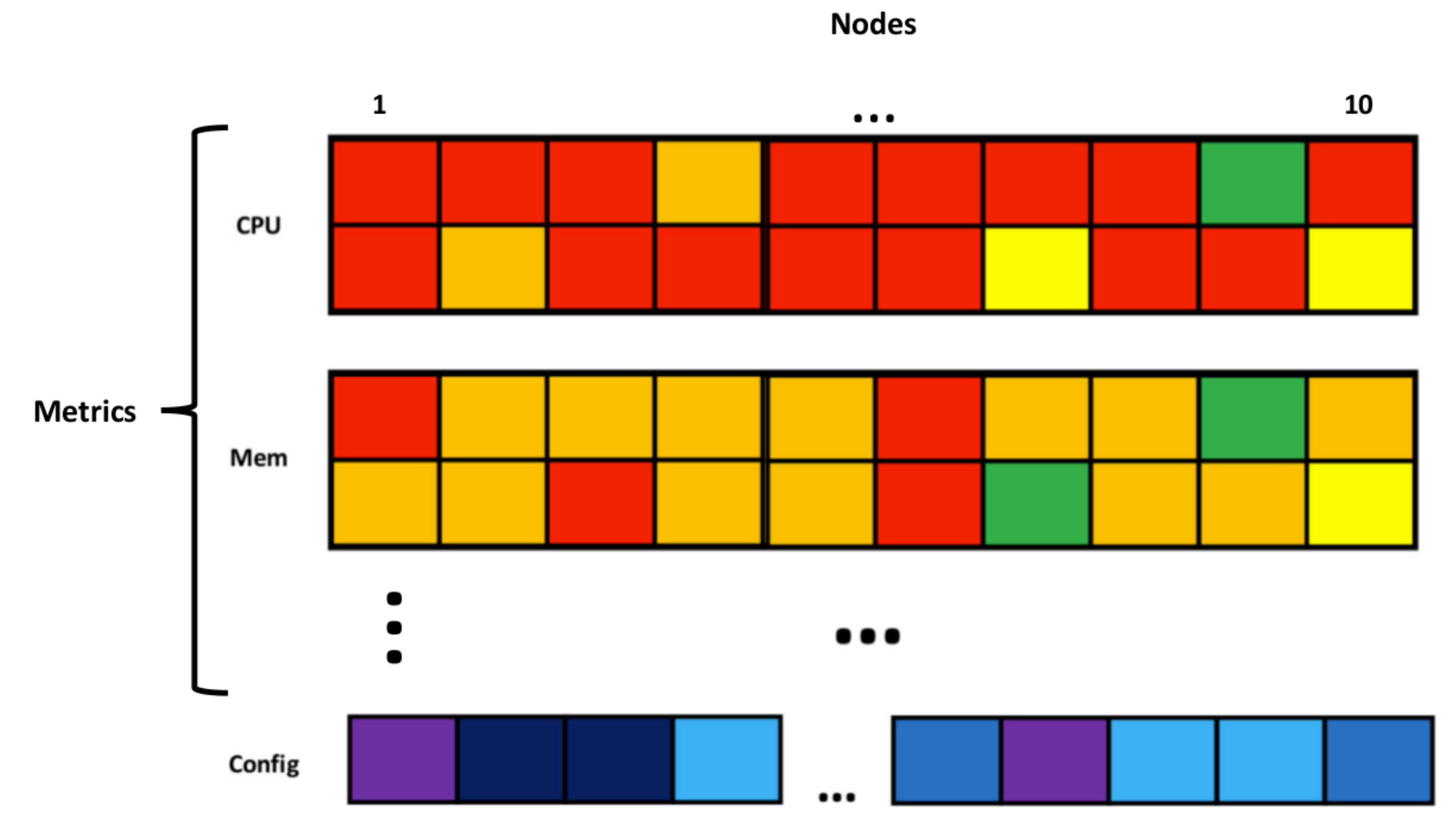}
\caption{An example of a state representation}
\label{fig:state}
\end{figure}

State (configuration plus metric values) could be represented as a heatmap of utilisation across nodes in the cluster. See Figure~\ref{fig:state} for a specific example of the input to the neural network.

%The action space (valid lever configurations) is kept small enough thanks to the selection process described above and also due to the system changing one lever at a time only. However, in order for the RL algorithm to explore new alternatives, we also permit the occasional selection of a different lever (see below).

We craft the reward signal to guide the agent towards good solutions for our objective: minimising event processing latency. For each tuning, we set the reward to $\sum_{e \in E} -1 / T_{e}$ where $T_{e}$ is the latency for event $e$ and $E$ is the set of events that can arrive to the streaming engine.

We represent the policy as a neural network (called policy network) which takes as input the collection of heatmaps described above, and approximates a probability distribution over all possible actions (as restricted by the output of the Lasso). Each tuning is based on a fixed number of events (with upcoming events being temporarily held until the phase completes).

The tuning phase terminates when all the events have been processed. Each tuning phase can apply several configurations $C$ (one change at a time) in order to reach an acceptable latency. A number of configurations $N < C$ within a tuning phase defines an episode. The value for $N$ is determined empirically.  We use $N\in(0.3C,0.8C)$. Higher values of $N$(closer to $C$) mean slower learning but a better chance of finding a more complex solution (or getting a worse performance). Lower values yield a more consistent behaviour.

%Empirically determined. We use N=(0.3C-0.8C). Higher (closer to C) means slower learning but a better chance of finding a more complex solution (or getting a worse performance). Higher is a bit like exploring (lack of reward for C - N means the algorithm drifts and behaviour is more oscillatory but you can find good solutions). Lower means more consistent behaviour.

Rewards are only applied at the end of each episode. We also set discount factor $\gamma$ to 1, so that the cumulative reward over time coincides with (negative) the sum of each event latency, hence maximising the cumulative reward mimics minimising the average latency.

State, action, and reward information for all configuration steps of each episode are recorded in order to compute the (discounted) cumulative reward, $v_{t}$, at each timestep $t$ of each configuration step and to train the neural network using a variant of the REINFORCE algorithm (shown in Algorithm~\ref{alg:reinforce}).

\begin{algorithm}
  \caption{Adapted REINFORCE algorithm, based on~\cite{Sutton1999, Mao2016}}
  \label{alg:reinforce}
  \begin{algorithmic}
    \While{ ( hasConverged!=true OR noMaxNumIter) }
    \State $\Delta \theta \gets 0$
    \For{$i=0$ to $N$}
      \State$(s_{1},a_{1},r_{1})...(s_{L_{i}},a_{L_{i}},r_{L_{i}}) \sim \pi_{\theta}$;
      \For{t : 1 .. $L_{i}$}
        \State$v_{t}^{i}=\sum_{s=t}^{L_{i}} \gamma^{s -  t} r_{s}^{i}$ \Comment{ compute returns}
      \EndFor
    \EndFor
    \For{t : 1 .. $L_{i}$}
      \State $b_{t} = \frac{1}{N} \sum_{1}^{N} v_{i}^{t}$ \Comment{compute baseline}
      \For{i : 1 .. N}
        \State$\Delta\theta \gets \Delta\theta + \alpha \nabla_{\theta} log \pi_{\theta} (s_{t}^{i},a_{t}^{i})(v_{t}^{i},b_{t}^{i})$
      \EndFor
    \EndFor
    \State$\theta \gets \theta + \Delta\theta$
    \EndWhile
    \State\textbf{return} $\theta$
  \end{algorithmic}
\end{algorithm}

To reduce the variance, it is common to subtract a baseline value from the returns, $v_{t}$. The baseline is calculated as the average of the return values, $v_{t}$, where the average is taken at the same configuration step across all episodes.

We employed a neural network with a fully connected hidden layer with 20 neurons. The heatmaps are blended to amalgamate 40 configuration steps and each episode lasts for 250 configuration steps. We used a similar setup to the one reported by~\cite{Chintapalli2016} (26 nodes to create 17000 events per second). We update the policy network parameters using the rmsprop~\cite{Mao2016} algorithm with a learning rate of 0.001. Unless otherwise specified, the results below are from training for 1000 training iterations.

\section{Experimental Evaluation}
\label{sec:eval}
To evaluate our work, we implemented our techniques using Google TensorFlow and Python's scikit-learn~\cite{Abadi2016, Pedregosa2007}. We loaded Spark with the well-known Yahoo streaming benchmark that simulates an advertisement analytics pipeline~\cite{Chintapalli2016}. We also tested the performance of the system on real-world production workloads for a top consumer IoT vendor company.

We conducted all of our deployment experiments on Amazon EC2.  We deployed our system controller together with the workload generator clients. These services are deployed on m4.large instances with 4 vCPUs and 16 GB RAM. The Spark streaming deployment was deployed over 10 m3.xlarge instances with 4 vCPUs and 15 GB RAM. We deployed our tuning manager and repository on a local server with 20 cores and 256 GB RAM.

We first perfomed a preliminary evaluation to determine:
\begin{itemize}
\item How long does it take to train the policy network?
\item Do changes make sense? How is latency affected?
\item Can we adapt to workload changes?
\item How does it compare to the performance obtained by two expert big data engineers for a production workload?
\end{itemize}

In the following subsections, we provide answers to all of these questions.

\subsection{Training Time}

\begin{figure}[t]
\epsfxsize=8.5cm \epsfbox{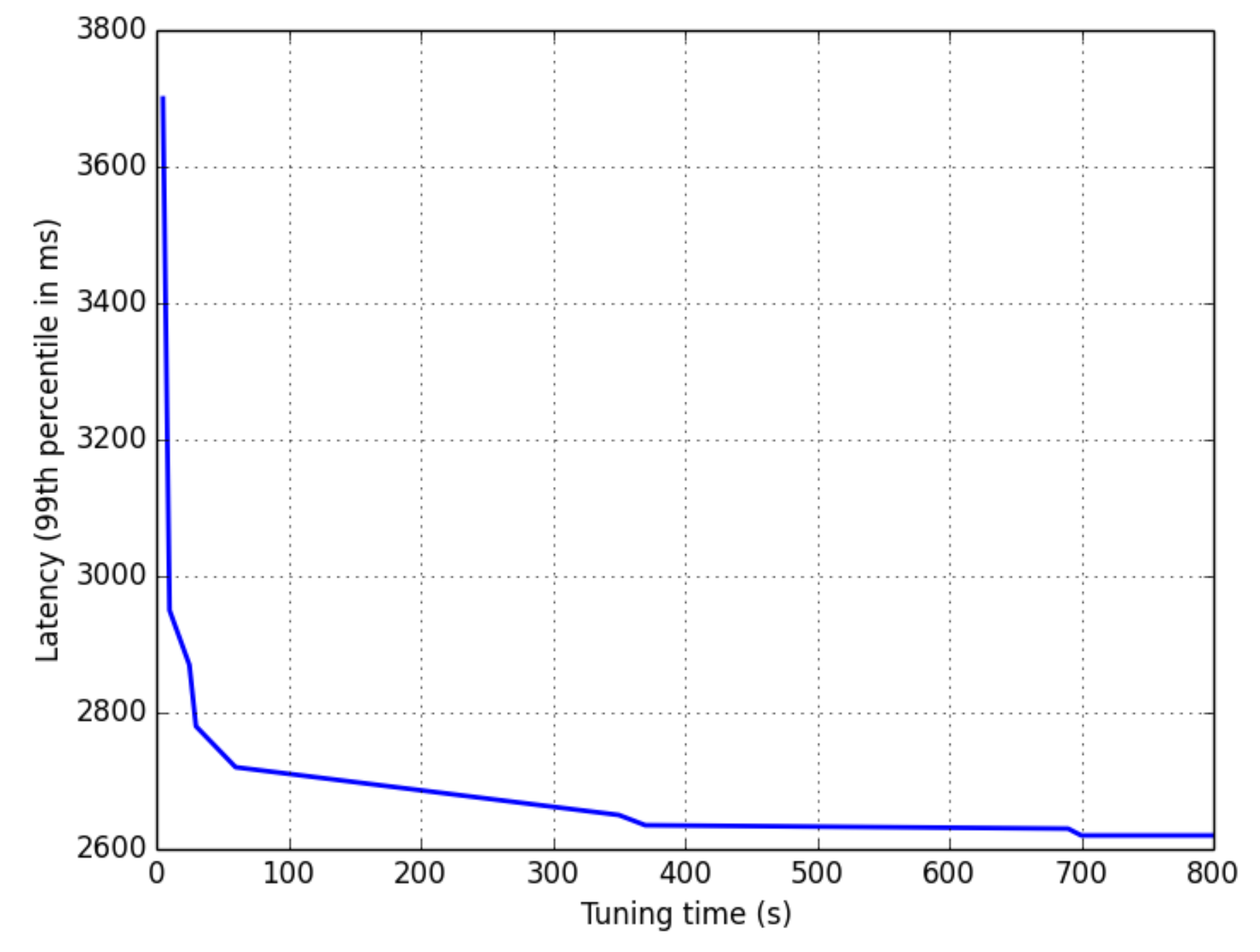}
\caption{Reduction in latency ($99^{th} percentile$) as training progresses.}
\label{fig:train}
\end{figure}

Every 5 minutes, the network tries a new configuration (changing just one lever at a time, $N=1$). Lasso is re-evaluated after each training phase, hence its output remains constant for the duration of the results shown in Figure~\ref{fig:train}. After 50 minutes, latency was reduced by more than 70\% ). 
Figure~\ref{fig:train} shows how latency decreases as the policy network selects new configuration values.
As can be observed in Figure~\ref{fig:train}, the first few iterations (departing from the default Spark configuration) are very productive in terms of latency reduction. Two of these changes were exploratory and resulted in no change in performance (some times transient performance decreases are observed during exploration). The remaining 7 changes were done on an exploitation mode. Three of these resulted in a minimum impact on performance. The training fully converged after 11h with increasingly smaller improvements. Better training times would have been achieved with GPU-boosted hardware, but reacting to changes in tens of minutes can be sufficient for a wide variety of streaming applications.

\subsection{Execution Breakdown}

To better understand what happens when computing a new configuration at the end of an episode, we logged the RL configuration output to record the amount of time spent in the different parts of the tuning process when running a workload similar to~\cite{Tibshirani2011}.

The workload is continuously executed but Kafka is buffering new incoming events during Configuration Loading and Preparation in case of node unavailability. This is possible since we designed our Spark jobs to behave idempotently by sinking their processed data on a set of partitioned tables.

The execution of an episode of the RL module can be broken down into:
\begin{itemize}
\item \textit{Configuration Generation}: time to calculate the best change to the current configuration.
\item \textit{Configuration Loading and Preparation}: time it takes to the system to install the new configuration and prepare Spark for the new tuning phase (incoming events being buffered by Kafka).
\item \textit{Workload Stabilisation}: we enable some time to enable some time for the changes to exert some effect in the workload. The stabilisation occurs before 3 min ($99\%$ of the time) but we dynamically detect stabilisation by creating trends on the variance of the latency and the most relevant metrics, as defined above.
\item \textit{Network Reward and Adaptation}: time to apply the reward and update the deep neural network parameters.
\end{itemize}

\begin{figure}[t]
\epsfxsize=8.5cm \epsfbox{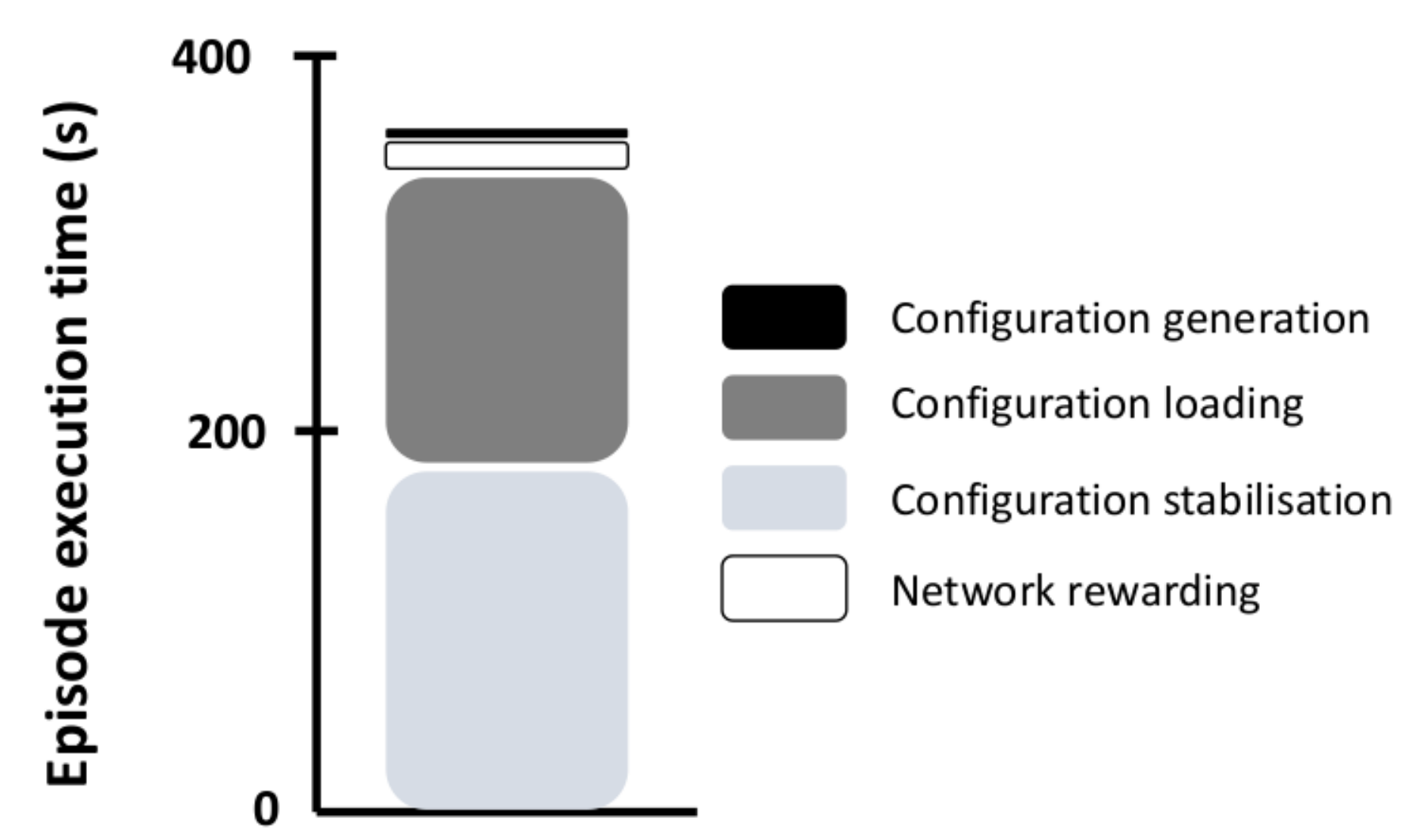}
\caption{Execution Time Breakdown – The average amount of time spent in the parts of the system during an episode.}
\label{fig:breakdown}
\end{figure}

The results in Figure~\ref{fig:breakdown} show the breakdown of the average times spend during an episode. As shown in the Figure, the time it takes to run an episode is dominated by two main factors: loading the new configuration and allowing for the configuration effects to reach a stationary state (a constant synthetic workload with 100K events throughput was used in this experiment). Depending on the suggested changes, the configuration loading is done without rebooting the nodes in the cluster, unless this is strictly needed for the configuration to take effect. The time it takes to apply the reward and update the network and to create the new configuration is negligible in comparison.

\subsection{Quality of the Suggested Changes}

\begin{figure}[t]
\epsfxsize=8.5cm \epsfbox{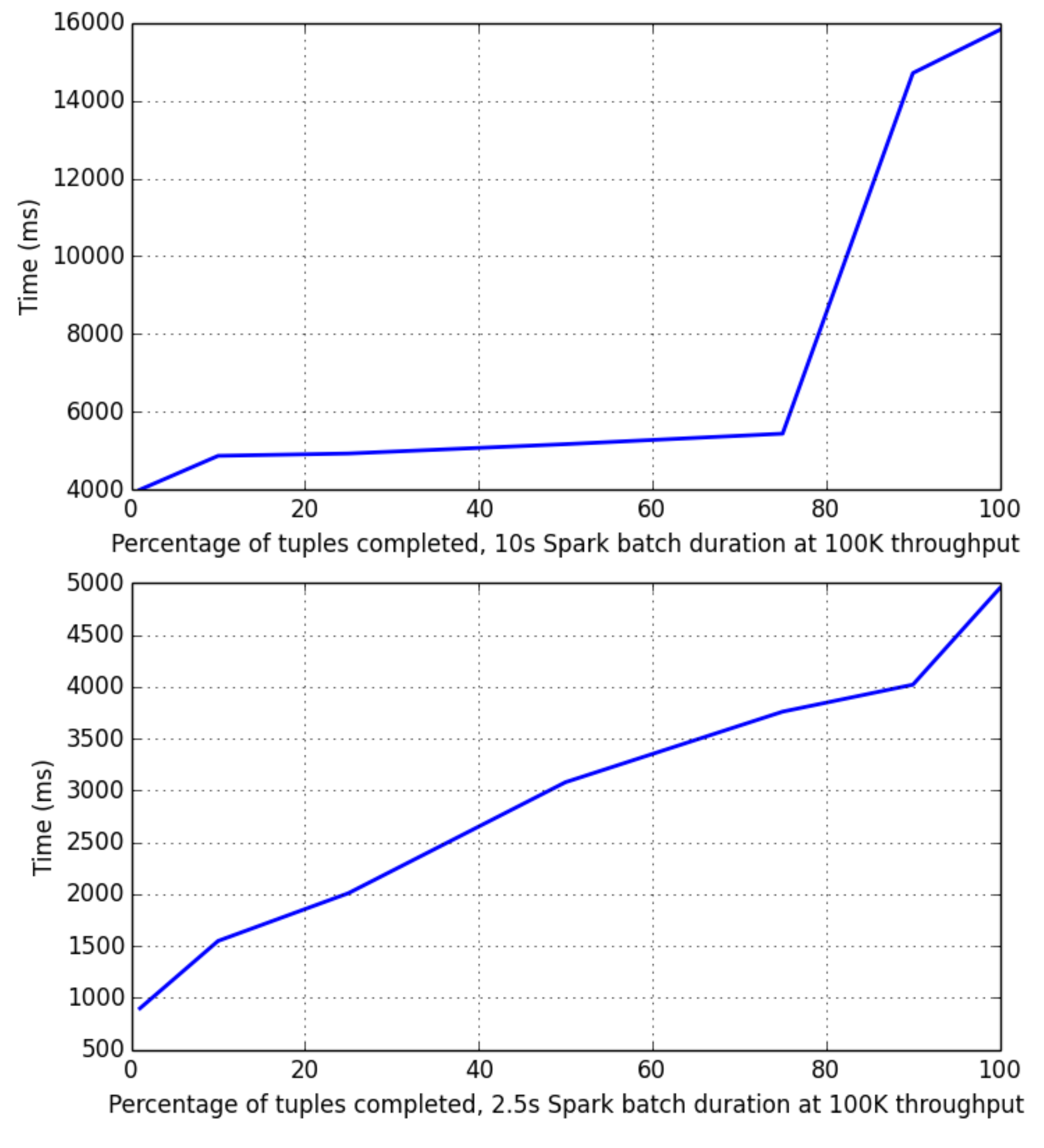}
\caption{CDF of the end to end latency ($99^{th} percentile$) for different configurations (\textit{Left:} 10s Spark batch size vs \textit{Right:} 2.5s Spark batch size).}
\label{fig:quality}
\end{figure}

We started with a batch setting interval of 10s, where the system can barely cope (and hence latency increases, see Figure~\ref{fig:quality}A. Then, the network automatically suggested to reduce the batch interval to 2.5s, resulting in a notorious improvement in latency at the highest throughout (Figure~\ref{fig:quality}B). This may seem to be a negligible difference to a human administrator, but the effects in performance are quite significant.

This is just an example of suggested configuration. The network starts with a default batch size of 10s. In our initial discretisation of the Spark batch size corresponded to the smallest bin that could be assigned (disregarding ripple effects). The dynamic discretisation mechanism described above enables the RL configurator to dynamically segment a bin into smaller 'sub-bins' in order to find the most appropriate configuration.

\subsection{Adaptation to Workload Changes}

In this subsection, we show how the RL configurator is capable of adapting to new radically different workloads. In this example, we use a synthetic benchmark that models a Poisson distribution in terms of event arrival rate.

$Y_{n}$ is the number of events entering the Kafka queue used by the Spark cluster to consume events during the $n^{th}$. We assume that for a short interval the rate of arrivals to this queue is fixed and that the distribution of this variable is Poisson, i.e. $Y_{n} \simeq Poisson(\lambda)$. In this case, we modelled two distributions $\lambda_{1}$ and $\lambda_{2}$ corresponding to arrival throughputs of 10000 and 100000, respectively.

We also model the size of the events for each of the two distributions above. We modelled two Gaussian distributions with similar standard deviation (0.3), but different means (0.5 and 5 MB, respectively). Thus, we have \textit{distribution 1}, which is low rate and small size-events, and \textit{distribution 2}, which is high rate and large-sized events.

\begin{figure}[t]
\epsfxsize=8.5cm \epsfbox{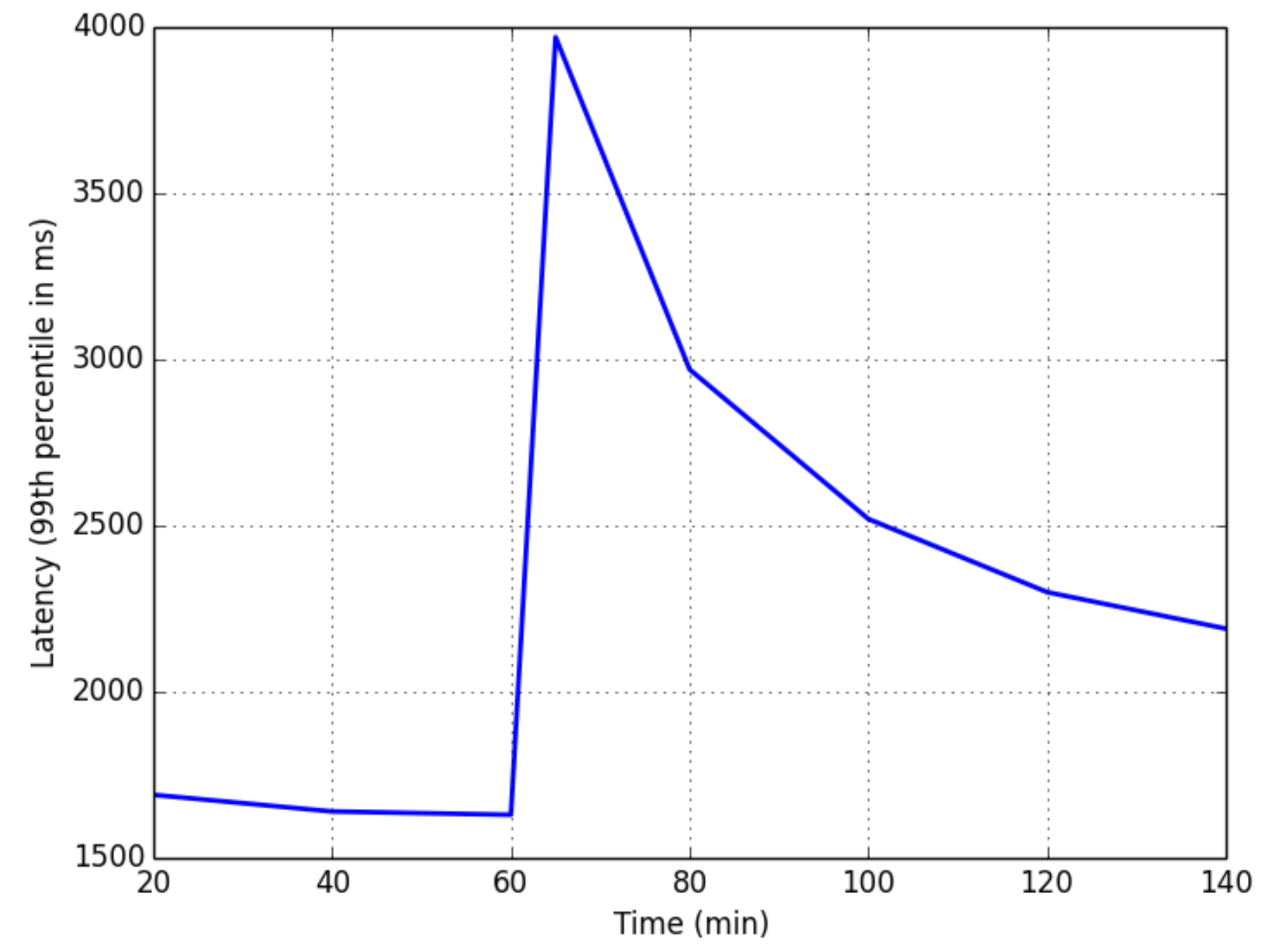}
\caption{Adaptation to drastic changes in workload.}
\label{fig:adapt}
\end{figure}

As can be observed in Figure~\ref{fig:adapt}, the workload is changed from distribution $\lambda_{1}$ to $\lambda_{2}$ around minute 65, resulting in a spike in the latency value that nearly doubles the previous baseline. The RL algorithm is capable of improving the situation but it cannot return to the previous baseline as larger events take longer to process \textit{distribution 2}.

\subsection{Exploration vs Exploitation}

 As described above, the best lever (according to our RL configurator) is used $f$ of the time. This is referred to as exploitation since the configurator ``exploits'' prior knowledge. This section explores the right value of $f$ depending on how frequently our workload changes.

 We alternate between distributions $\lambda_{1}$ and $\lambda_{2}$ 1, 3 o 6 times per hour. We then measure the time it takes for the RL configurator to reach 80\% of its previous baseline value. By baseline, we mean the stationary latency that is reached when neither workload nor resources are changed. These values (baselines) are $\simeq 2000ms$ and $\simeq 3200ms$ for \textit{distribution 1} and \textit{distribution 2} as shown in Figure~\ref{fig:adapt}.

 \begin{table}[]
 \centering
\begin{tabular}{|c|cc|cc|cc|}
    \hline
    \diagbox{$f$}{rate} & \multicolumn{2}{c|}{\textbf{1/60}}  & 
    \multicolumn{2}{c|}{\textbf{3/60}}   & \multicolumn{2}{c|}{\textbf{6/60}}  \\
    \hline 
    0.9 & 18.9 min & \textit{1} & 19.1 min & \textit{1.26} & 10 min & \textit{1.5} \\
    0.8 & 18.1 min & \textit{1} & 18.8 min & \textit{1.12} & 10 min & \textit{1.4} \\
    0.7 & 16.5 min & \textit{1} & 17.1 min & \textit{1.05} & 10 min & \textit{1.2} \\
    \hline
    
\end{tabular}
 \caption{Convergence times and baseline (italics) for different values of the Exploration vs. exploitation factor $f$ (rows) at different rates of workload change (columns).}
 \label{tab:exp}
 \end{table}

In Table~\ref{tab:exp}, we show the time to reach a stationary value (top of each cell) and the level above the initial baseline that the RL configurator was capable of obtaining. As frequency increases, the RL configurator does not have enough time to find the right configuration and the experiment terminates with a latency value that is higher than the reported baseline.

As can be observed in the top number within each cell in the table, more exploration (lower $f$) implies faster adaptation (measured as time to reach 1.2 times the original baseline) for changes in workload even at higher frequencies.

Higher values of $f$ result in worse baselines (bottom number of each cell in the table) for the same frequencies since the RL configurator has a more restricted ability to explore new configurations. The downside of lower $f$ values is increased variability (standard deviation in the mean values reported in Table~\ref{tab:exp} are 0.15, 0.26, and 0.34 for $f=0.9$, $f=0.8$, and $f=0.7$, respectively).

\subsection{Comparison to Human Configurators}

We tested how different mechanisms can be employed to configure the network. The results in these section are not meant to be exhaustive and should be taken as a qualitative indication only.

We took 2 expert data engineers with more than 10 years of industrial experience and gave them 1 day to do their tweaks in the cluster configuration. We also recruited 9 students from a Computer Science MSc. All of the students had previously taken a unit with several lectures and assignments on how to configure Spark. Students were given a full week to deliver their best configuration. We compared these two cohorts with our algorithm.

\begin{figure}[t]
\epsfxsize=8cm \epsfbox{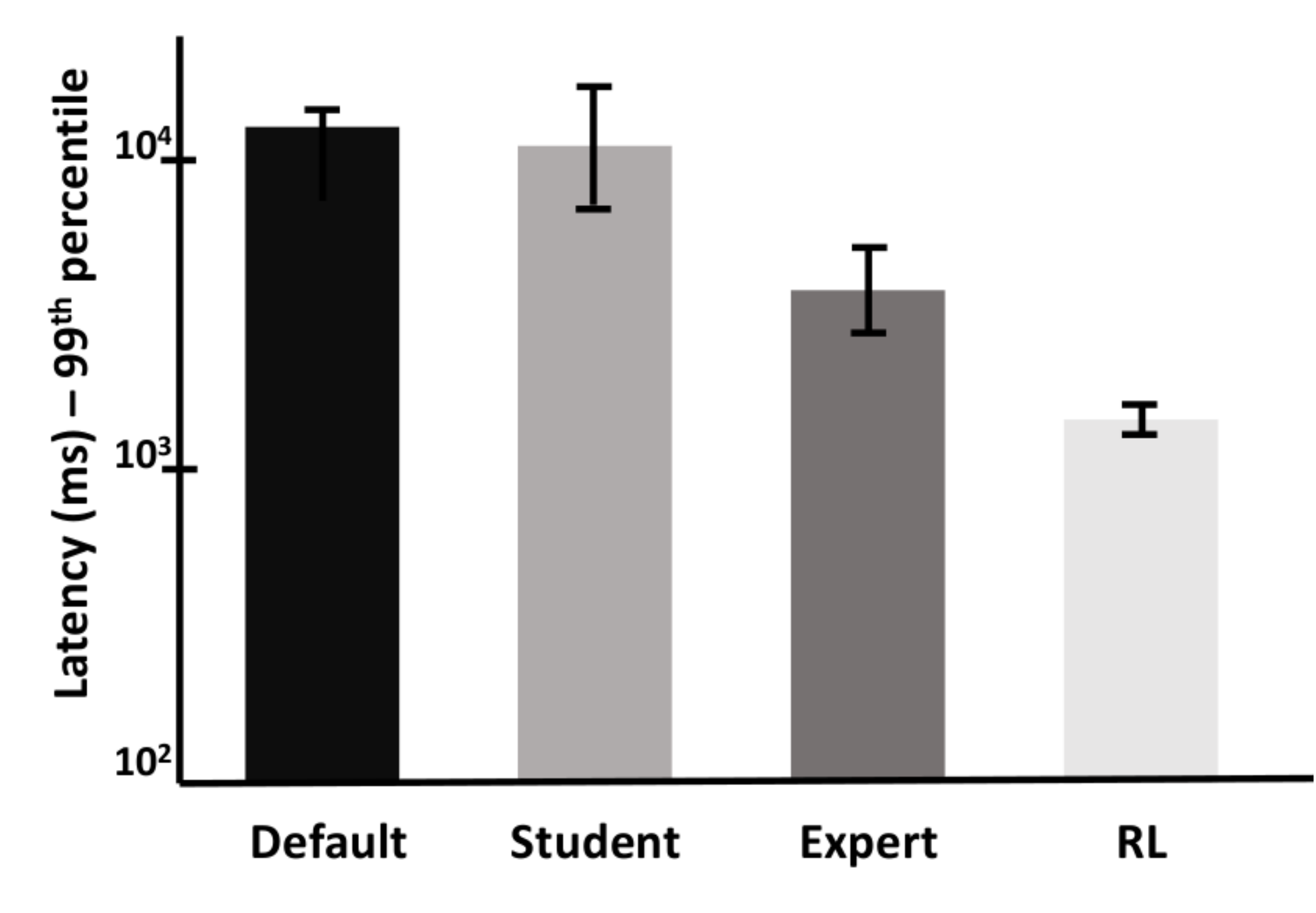}
\caption{Comparison of the results obtained in tuning the cluster by several different methods.}
\label{fig:humanperf}
\end{figure}

As can be observed in Figure~\ref{fig:humanperf}, the RL network is more efficient than their human counterparts. Unsuprisingly, the experts seemed to be better than students, but the small samples size here prevents us from making any strong claims.

These results were obtained with differences in the time it took each method to accomplish the reduction in latency. The RL method is capable of reaching much better configurations in a fraction of the time it takes humans. Note that the results performed for the RL method are the ones obtained after 50 min of running. As previously shown in Figure~\ref{fig:train}, further improvements would have been possible just by letting the RL run for 10h (still significantly less than their human counterparts).

%\begin{figure}[t]
%\epsfxsize=8cm \epsfbox{fig10.pdf}
%\caption{Comparison of the time employed by each method to tune the cluster configuration.}
%\label{fig:humantime}
%\end{figure}

\section{Related Work}
\label{sec:rw}

\subsection{Configuration Sampling}

Performance prediction techniques can: 1) compile all possible configurations and record the associated performance scores (maximal sampling), which can be impractically slow and overly expensive and time consuming~\cite{Weiss2008}; or 2) intelligently selecting and executing ``enough'' configurations to build a predictive model (minimal sampling). For example, Zhang et al.~\cite{Zhang2015} approximate the configuration space as a Fourier series to derive an expression showing how many configurations must be studied to build predictive models with a given error. Continuous learning techniques have also be applied to ensure adaptability~\cite{Thalheim2017}. Our work falls closer to this later set of continuous learning techniques, but it also relies on gathering information on a large number of samples (minimal sampling) as an exhaustive screening of the full configuration space is simply unfeasible.

One of the problems with massive configuration spaces is derived from the fact that many configuration parameters are continuous in nature and can, hence, take an infinite number of values. We built on previous work~\cite{Subiros2015} to dynamically discretise continuous configuration variables.

\subsection{Metric Dimensionality Reduction Techniques}

Modern monitoring frameworks have created an opportunity to capture many aspects of a the performance of a system and the virtualised environment it tends to run on. This has resulted in an information crosstalk and overload problem where many metrics are non-linearly interdependent.

Reducing the size and dimensionality of the bulk of metric data exposed by complex distributed systems is essential. Common techniques include sampling to enable a systematic trade-off between the accuracy, and efficiency to collect and compute on the metrics~\cite{Zhou2000, Kollios2003, Krishnan2016, Quoc2017}, and data clustering via k-means and k-medoids~\cite{Ng1994, Ding2015}.

Classic approaches such as principal component analysis (PCA)~\cite{pearson1901lap} and random projections~\cite{Papadimitriou1998} can also be used for dimensionality reduction, these approaches either produce results that are not easily interpreted by developers (i.e., PCA) or sacrifice accuracy to achieve performance, producing different results across runs (i.e., random projections). On the other hand, clustering results can be visually inspected by developers, who can also use any application-level knowledge to validate their correctness. Additionally, clustering can also uncover hidden relationships which might not have been obvious.

\cite{Thalheim2017} et al., focus on analysing interdependencies across metrics by building a call graph. Similar to~\cite{VanAken2017}, we rely on factor analysis to determine the most relevant metrics, hence reducing the problem of metric dimensionality. Like these authors, we also rely on Lasso to find the strongest associations between metrics and configuration levers.

\subsection{Machine Learning for System Configuration Optimisation}

The usage of machine learning methods to tweak configuration in systems is not novel. For instance,~\cite{Zhang2018} review on the usage of deep learning in networking configuration.

Previous work on self-tuning databases is focused on standalone tools that target only a single aspect of the database, such as indexes~\cite{Skelley2000} or partitioning schemes~\cite{Agrawal2004}. Other tools are workload-specific~\cite{Debnath2008}. They all require laborious preparation of benchmark workloads, spare hardware and expertise on the database internals, which~\cite{VanAken2017} does not require. OtterTune uses GP regression to learn workload mappings~\cite{VanAken2017}). GP offers several advantages like the dynamic tradeoff between exploration and exploitation, but it relies on an explicit process of performance prediction.

More recent efforts have focused on optimising the configuration of in memory databases~\cite{pavlo17}.

STREAM~\cite{Arasu2003}, Aurora~\cite{Balakrishnan2004} and Borealis~\cite{abadi2005} were the precursors of a Cambrian explosion in the variety of streaming engines~\cite{Akidau2013, Kulkarni2015, Neumeyer2010} many of which like (Twitter’s Heron, Storm, Samza, Flink or Spark Streaming) have been open-sourced. Despite all their sophistication and performance none of the existing streaming systems are truly self-regulating.

\cite{Floratou2017} presented an architecture enabling streaming engines to self-regulate. They presented policies to adjust the topology configuration so that the performance objectives are met even in the presence of slow machines/containers, similar to~\cite{Fu2015,Gedik2014} but lacking the ability to automatically scale based on the input load.

\cite{Herodotou11} presents self-tuning techniques for Map Reduce systems based on a graph of workload that can be optimised.~\cite{Dalibard2017} applied Bayesian optimisation techniques to garbage collection. Recent work proposed self-driving relational database systems that predict future workloads and proactively adjust the database physical design~\cite{pavlo17}.

In recent years, Deep reinforcement learning (DRL) has gained great success in several application domains. Early works use RL for decentralised packet routing in a switch at small scales~\cite{Boyan1993}. Congestion protocols have also been optimised online and offline using RL~\cite{Dong2015, Winstein2013}. Cluster scheduling has been studied widely recently~\cite{Chen2017, Grandl2014, Isard2009, Mao2016, Zaharia2010}.

Unlike prior efforts, our system does not focus on topology configuration, job scheduling or routing confiuration, but on finding which configuration parameters in a streaming engine (Spark Streaming in our case) make a difference to maintain predefined latency/throughput SLOs.

\subsection{Machine Learning for Workload Prediction}

Large configuration spaces are a common theme in the literature. As mentioned above, sampling has been a the gold standard to try to tackle this problem ~\cite{Siegmund2012, Guo2013, Sarkar2015}. These solutions tend to require manual configuration, while subjecting the learning systems to very large variance~\cite{Nair2017}. For instance, regression tree techniques for performance prediction require thousands of specific system configurations~\cite{Guo2013}, even when the authors used a progressive random sampling approach, which samples the configuration space in steps of the number of features of the software system in question. Sarkar et al.~\cite{Sarkar2015} randomly sampled configurations and used a heuristic based on feature frequencies as termination criterion. The samples are then used to train a regression tree.~\cite{Nair2017} used eigenvalues of the distance matrix between the configurations to perform dimensionality reduction to minimise configuration sampling by dropping out close configurations while measuring only a few samples.

While these are related to our approach, we do not intend to build a performance predictor based on metrics and configurator. Our approach uses several intermmediate steps: 1) selecting relevant metrics, 2) associating metrics to right configuration levers, and 3) learning association of metric to lever in order to improve the performance of the system. Our system uses techniques similar to~\cite{VanAken2017} for steps (1) and (2). In our system learning is confined to the third phase, which requires no prior configuration sampling.

Gaussian Process (GP) Models (GPM) is often the surrogate model of choice in the machine learning literature. GPM is a probabilistic regression model which instead of returning a scalar ($f(x)$) returns the mean and variance associated with $x$. Building GPMs can be very challenging since they can be very sensitive to the parameters of GPMs, they do not scale to high dimensional data as well as a large dataset (software system with large configuration space)~\cite{Shen2005} and can be limited to models with around ten decisions~\cite{Wang2016}.

\section{Discussion}
\label{sec:disc}

Our system explores this massive configuration space by using Reinforcement Learning. We have shown how the system selects obvious configuration levers in the first few episodes (e.g. increasing the memory of the driver node), resulting in substantial performance gains. The system can be tweaked with a single parameter ($f$), allowing data engineers to balance between configuration exploration and exploitation. Higher exploration rates have been found to obtain better solutions, although the higher variance increases the likelihood of ``faulty configurations'' (ones where the cluster cannot keep running). Data engineers do not need to take care of machine boundaries (e.g. configuring worker nodes consistently), as the system does this on their behalf.

%I don't undestand what is the messagehere
%Our work is not constrained by machine boundaries and locality. Mao et al.~\cite{Mao2016} abstract away from machine boundaries and potential resource fragmentation by assuming a single resource pool. Their approach works since many cluster schedulers make independent scheduling decisions per machine (such as YARN~\cite{Vavilapalli2013}). These authors use RL for job scheduling, which is different to consistently applying configurations across a set of distributed nodes.

The learned behavior, including the right balance for ($f$), is specific to the workload, job/analysis, and cluster type, requiring an abundance of data to tweak the RL configurator. As future work, we plan to explore transfer learning techniques to minimise this need~\cite{Yang2017,Jamshidi2017}, opening the applicability of this technique to more heterogeneous scenarios.

Stream processing systems can be fine tuned to accommodate different workloads in a variety of ways. However, the ability to remain performant under changing workloads is a fundamental aspect. A range of techniques have been suggested to address this challenge, including scaling the number of virtual machines used in the streaming engine~\cite{Cervino2012}, using smarter mechanisms to allocate workloads to dynamic cores~\cite{Micolet2016} or cluster nodes~\cite{Mao2016}, dynamic load balancing~\cite{Martins2014}, or even specific configuration aspects (like batch size~\cite{Das2014}). We have shown in this paper how our approach can react automatically to configuration changes, and preserve low streaming latencies by automated learning and exploration of the vast configuration space.

Our system is based on a set of algorithms that operate optimally under clear domain restrictions. The effectiveness of our system depends on having a linear relationship between central metrics and levers~\cite{Tibshirani2011}. We found these to apply in our experiments, but that might not be the case in other systems with different behavior, configuration levers and obtained metrics.

We have taken advantage of the lower intrinsic dimensionality of configuration spaces~\cite{Nair2017}, by using random sampling techniques to help reducing the dimensionality of our model. This approach allowed us to implement RL methods in a problem where it would seem to be unfeasible.

%which reduces the effective cost of modeling  While it is seemingly As pointed out by~\cite{Nair2017}, problems with lower dimensions should be easier/cheaper to model than those with higher intrinsic dimensionality. Nair et al. explore the lower intrinsic dimensionality of configuration spaces, and exploit those lower dimensions for the purposes of sampling. In our work, we use random sampling and then try to reduce the number of dimensions on a large data set.

%I don't think this one either is worth mentioning; we mostly cite another paper, and
%Finding a near-optimal configuration can become challenging when the configuration space is non-convex, as required by gradient descent methods. In our evaluation, we could not find this (similar to~\cite{Nair2017}). Our system does not find the optimal configuration (gradient descent can be trapped in a local minimum, some dependencies may be supra linear but still, etc) but a configuration that can improve and adapt over time in an automated manner.

Our system uses a close time horizon (to compute the baseline in Algorithm~\ref{alg:reinforce}) whereas the underlying optimisation problem has a far time horizon (data streams are infinite in theory). Value networks estimating average return values can be used to overcome this limitation in the future~\cite{Sutton1998}.

The main overhead of our system is the machine reboot time after a new configuration As part of our future work, we want to explore how the system performs by restricting its behaviour to those configurations that do not require rebooting the cluster. Another potential approach to minimise downtimes is a green-blue deployment setting where changes are applied into a new cluster. Both, the blue and green cluster read events from our Kafka topics and idempotently dump the data into a database. Tuning the time to entirely move the workload to the new cluster requires job dependent techniques that we are starting to explore.

\section{Conclusions}
\label{sec:conc}

We presented the first stream processing system that uses RL to adapt to a variety of workloads in a dynamic manner. Our system converges to better solutions than human operators in much less time, resulting in significant latency (60-70\%) reductions in a few tens of minutes. The system adapts well to different workloads while requiring minimal human intervention.

\section{Acknowledgments}

{\footnotesize \bibliographystyle{acm}
\bibliography{mw2018}
}

%\theendnotes

\end{document}